\documentclass[a4paper,11pt]{article}
\usepackage{de}
\usepackage{orcidlink}

\usepackage{wrapfig}
\title{Exploring the physics of ram pressure stripping with radio continuum observations in the SKA era}
\ShortTitle{Exploring the physics of ram pressure stripping in the SKA era}

\author[1]{Alessandro Ignesti \orcidlink{0000-0003-1581-0092}}
\ShortName{Alessandro Ignesti et al.} 

\author[2,3]{Ian D. Roberts \orcidlink{0000-0002-0692-0911}}
\author[4,5]{Henrik W. Edler \orcidlink{0000-0002-4526-4806}}
\author[6,7]{Myriam Gitti \orcidlink{0000-0002-0843-3009}}
\author[8]{Reinout J. van Weeren \orcidlink{0000-0002-0587-1660}}
\author[9]{Paolo Serra \orcidlink{0000-0001-5965-252X}}
\author[10]{Pavel Jachym \orcidlink{0000-0002-1640-5657}}

\affiliation[1]{INAF - Astronomical Observatory of Padova, vicolo dell'Osservatorio 5, 35122 Padova, Italy }
\emailAdd{alessandro.ignesti@inaf.it}
\affiliation[2]{Waterloo Centre for Astrophysics, University of Waterloo, 200 University Avenue West, Waterloo, ON, N2L 3G1, Canada}
\affiliation[3]{Department of Physics $\&$ Astronomy, University of Waterloo, 200 University Avenue West, Waterloo, ON, N2L 3G1, Canada}
\affiliation[4]{ASTRON, Netherlands Institute for Radio Astronomy, Oude Hoogeveensedijk 4, 7991 PD Dwingeloo, The Netherlands}
\affiliation[5]{Hamburger Sternwarte, University of Hamburg, Gojenbergsweg 112, D-21029, Hamburg, Germany}
\affiliation[6]{Dipartimento di Fisica e Astronomia (DIFA), Universit{\`a} di Bologna, via Gobetti 93/2, 40129 Bologna, Italy}
\affiliation[7]{ INAF - Istituto di Radioastronomia, via P. Gobetti 101, Bologna, Italy}
\affiliation[8]{Leiden Observatory, Leiden University, PO Box 9513, 2300 RA Leiden, The Netherlands}
\affiliation[9]{INAF,Osservatorio Astronomico di Cagliari, Via della Scienza 5, Selargius, 09047, CA, Italy}
\affiliation[10]{Academy of Sciences, Bo\v cn\'i II 1401, 141 00, Prague, Czech Republic}

\abstract{Satellite galaxies in clusters are significantly more likely to be red and passive than similar mass galaxies in the field. This fact is known as the environmental quenching of galaxy star formation, which is believed to be driven by ram pressure stripping (RPS). The large velocity differences between the infalling galaxies and the intracluster medium (ICM) result in a strong ram pressure on their interstellar medium (ISM), which can strip it from the stellar disk. The stripped ISM can be studied at various wavelengths, including the radio band, thanks to the synchrotron emission produced by the magnetic fields and relativistic electrons embedded in them. This emission is typically steep-spectrum and thus best observed at low frequencies. Thus, continuum studies of the RPS effect are currently mostly carried out with LOFAR, limiting them to the northern hemisphere. SKA-Low will permit us to extend them to the southern sky, where they will synergize with the southern observatories and the upcoming ELT. Lastly, the sub-arcsecond resolution provided by SKA-Mid will facilitate the exploration of the polarization and filamentary structure of RPS radio tails and allow us to detect them up to $z\simeq0.5$, advancing our understanding of the impact of RPS on satellite galaxies in clusters and groups.}
\begin{document}
\maketitle
\section{Environmentally-driven galaxy evolution}
The balance between the gas inflow and removal drives the baryon cycles in galaxies, and thus their evolution. Whereas the inflows sustain star formation and allow galaxies to grow in mass, gas removal can lead to star formation quenching. Galaxies can lose their gas via internal or external processes. The latter, generally known as environmental effects,  are crucial in shaping galaxy properties in clusters and groups \citep[][]{Dressler1980,Vulcani_2022,Boselli_2022}.
\begin{wrapfigure}{r}{0.5\textwidth}
  \begin{center}
    \includegraphics[width=\linewidth]{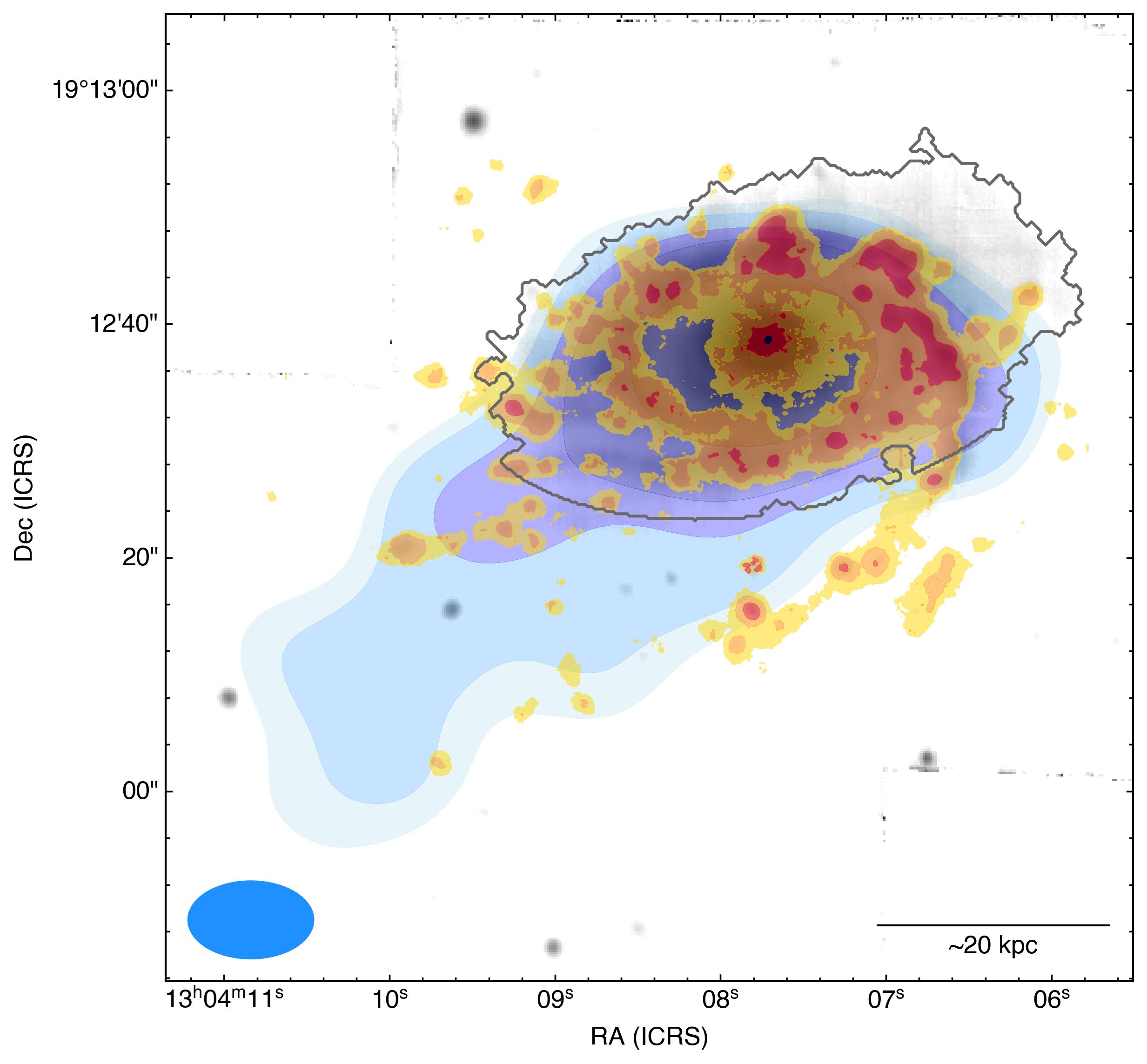}
  \end{center}
  \caption{\label{figura}Composite optical-radio image of the RPS galaxy JW39 from \cite{Ignesti_2022d}. Here are reported the stellar continuum (greyscale), stellar disk (silver contour), H$\alpha$ emission (yellow-to-red contours) and radio continuum emission at 144 MHz (blue-filled contours). }
\end{wrapfigure}
They can be divided into two main categories, those driven by gravitational forces between galaxies, such as mergers and tidal interaction \citep[][]{Barnes_1992} or harassment due to fast encounters \citep[][]{Moore1996}, and those resulting from the hydrodynamical interaction between the galaxies and the environmental plasma, either the intracluster medium (ICM) or the intragroup medium (IGrM). This category includes thermal evaporation \citep[][]{McKee-Cowie_1977}, viscous stripping \citep[][]{Nulsen1986}, and ram pressure stripping \citep[RPS,][]{Gunn1972}. The latter is the external pressure exerted by the environmental plasma on a moving body, typically expressed as $P_{\text{ram}}=\rho V^2$ where $\rho$ is the medium density and $V$ is the object velocity. In the most extreme cases, ram pressure can overcome the stellar disk binding force and strip the circumgalactic and interstellar medium (ISM) components from the galaxy \citep[][]{Gunn1972}. The gas loss induced by RPS can effectively quench the star formation in the stellar disk \citep[][for a review]{Boselli_2022}, thus making it an important quenching pathway for satellite galaxies \citep[e.g.,][]{Vollmer_2001,Tonnesen_2007,Vulcani2020,Watts_2023}. Although RPS is expected to be the strongest during the galaxies' first infall, especially for galaxies following the most radial orbits \citep[][]{Biviano_2024}, optical studies revealed that almost all galaxies in clusters undergo an RPS event within their lifetime \citep[][]{Vulcani_2022}. Moreover, due to the difference in the average environmental plasma density and satellite velocities, RPS is expected to dominate the galaxy evolution in clusters, where due to the large masses ($>10^{14}~M_\odot$ ) galaxies have a typical velocity dispersions of several hundreds to thousand of kilometers per second and the medium particle density is of the order of $10^{-4}-10^{-3}$ cm$^{-3}$, but thought to be less relevant in galaxy groups, lower mass systems ($<10^{14}~M_\odot$) where gravitational interactions are more frequent due to the generally smaller velocity dispersion. The ram pressure action is not limited to displacing the ISM outside of the disk. The impact of ram pressure can result in many effects, including compression of gas along the leading edge of the disk \citep[e.g.,][]{Rasmussen_2006,Poggianti_2019,Roberts_2022e}, disturbed galaxy morphologies,  trailing tails of stripped gas \citep[e.g.,][]{Kenney2004,vanGorkom_2004,Fumagalli2014,Poggianti2017}, and condensation of star-forming knots in the tails \citep[][]{Kenney2014,Poggianti2019}. It can also temporarily enhance the global star formation \citep[e.g.,][]{Poggianti_2016,Vulcani2018,Roberts_2020} and trigger the activity of the central nuclei \citep[e.g.,][]{Poggianti2017b,Peluso_2021}. Indeed it has been observed that it can affect the microphysics of the ISM on small scales, for example by stimulating the conversion from atomic to molecular hydrogen \citep[][]{Moretti_2020}, enhancing the diffusivity of cosmic rays \citep[][]{Farber_2022,Ignesti_2022d}, or by inducing mixing between the ISM and ICM \citep[][]{Sun_2022,Franchetto_2021}.
The most extreme examples of galaxies undergoing strong RPS are the so-called jellyfish galaxies \citep{Fumagalli2014, Smith2010, Ebeling2014,Poggianti2017}. In the optical/UV band, these objects show extraplanar, unilateral debris extending beyond their stellar disks, and striking tails of ionised gas hosting star-forming regions (Figure \ref{figura}). Jellyfish galaxies mostly reside in galaxy clusters and are a transitional phase between infalling star-forming spirals and quenched cluster galaxies; for this reason, they provide a unique opportunity to understand the impact of gas-removal processes on galaxy evolution. 
\section{Exploring the physics of ram pressure stripping with radio continuum observations}
\begin{figure}
    \centering
    \includegraphics[width=\linewidth]{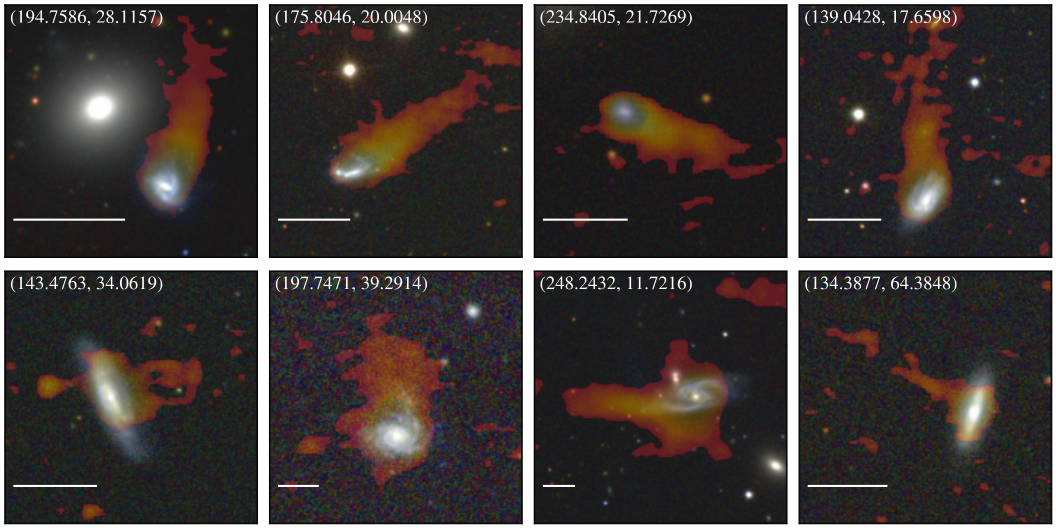}
    \caption{Composite optical (CFHT, RGB) and radio (LOFAR, 144 MHz) images of eight LoTSS jellyfish galaxies from \cite{Roberts_2021a,Roberts_2021b} at $z<0.04$. The galaxy celestial coordinates are reported in the top-left corner, and the white scalebar corresponds to a size of 20 kpc at the hosting cluster redshift.}
    \label{fig_LOFAR}
\end{figure}

A number of ram pressure stripped galaxies have been observed to show tails of radio continuum emission extending for tens of kiloparsecs from their stellar disk \citep[e.g.,][]{GavazziJaffe1987,Murphy2009,Vollmer_2013,Chen_2020, Muller_2021,Roberts_2021b,Roberts_2021c,Ignesti_2022d}. These radio tails typically develop steep-spectrum emission ($\alpha<-0.9$ at GHz frequencies) within few tens of kiloparsecs from the stellar disk \citep[e.g.,][]{Vollmer_2004,Chen_2020,Ignesti_2021,Muller_2021,Lal_2022,Venturi_2022,Roberts_2024a}. Therefore, they are best observed below GHz frequencies, and for this reason, the LOw Frequency ARray \citep[LOFAR,][]{vanHaarlem_2013} has been revolutionary for the study of RPS in clusters and groups. Specifically, the LOFAR Two-metre Sky Survey \citep[LoTSS,][]{Shimwell_2017,Shimwell_2019,Shimwell_2022} has provided high-resolution ($6 ``$) and highly sensitive ($\sim100$ $\mu$Jy beam$^{-1}$) images of the Northern sky at 120-168 MHz, revealing that at low frequencies, the stripped tails can extend for up to 100 kpc in the galaxies’ wakes, making them easily detectable in the radio sky (Figure \ref{fig_LOFAR}). Thanks to LOFAR, over one hundred new RPS galaxies have been discovered in the northern sky with an occurrence rate varying from 10 to 20$\%$ \citep{Roberts_2021b}. This indicates that RPS is not an isolated occurrence but rather a common phase experienced by the majority of cluster galaxies. With this relatively large sample size, it is now possible to conduct statistical studies of the RPS radio continuum tail development in clusters \citep[see][for a recent example]{Smith_2022}. 

RPS deeply affects the galaxies' nonthermal radio emission, from the microphysics regulating the ISM up to the overall morphology. In general, radio continuum emission in spiral galaxies is composed of the thermal emission from the $\sim10^4$ K plasma in the HII regions produced via thermal bremsstrahlung, and the non-thermal synchrotron emission of the relativistic cosmic rays electrons (CRe), which are accelerated by supernovae shocks reaching energies of a few GeV \citep[e.g.][for a review]{Condon_1992}. Given the connection between star-formation and CRe and the fact that galaxies are generally optically-thin to radio wavelengths, the continuum radio emission is a reliable proxy to evaluate their star formation rate (SFR) \citep[e.g.,][]{Kennicutt_2012}. At GeV energies, CRe energy losses are dominated by synchrotron emission in the galactic magnetic fields with typical values of 10-30 $\mu$G \citep[e.g.,][and references therein]{Beck_2005}, which results in nonthermal radio emission in the 100 MHz-10 GHz band, and Inverse Compton scattering with the galactic and stellar radiation field  \citep[e.g.,][]{Pacholczyk_1970,Longair_2011}. Star-forming galaxies in clusters usually show a large scatter in the radio luminosity-SFR relation with respect to isolated galaxies \citep{Chen_2020}, whereas RPS galaxies are typically in excess of radio luminosity with respect to their ongoing star formation rate. The excess can either be real, induced by the amplification of the ISM magnetic field by turbulence and compression, or apparent, resulting from a mismatch between the declining star formation rate and the CRe radiative time-scale \citep{Ignesti_2022d,Edler_2024}. Finally, by comparing the resolved spatial correlation between H$\alpha$ emission, which traces the ionized gas and, hence, the star-forming regions, and low-frequency radio continuum, which maps the nonthermal ISM, it has been observed that that the typical CRe transport scale in these extreme galaxies is 5-10 kpc, which is larger than the average values of 1-5 kpc measured in nearby, isolated systems \citep[][]{Ignesti_2022d,Edler_2024}. 

RPS is also responsible for the formation of the radio continuum tails, produced by the CRe accelerated in the stellar disk \citep[][]{Condon_1992}. Multifrequency studies observed that spectral index steepens with radial distance along these tails \citep[e.g.,][]{Vollmer_2004,Chen_2020,Muller_2021,Ignesti_2021,Roberts_2021c,Venturi_2022}, suggesting the following scenario. After the injection, the relativistic electrons are stripped, together with the ISM, from the disk by the ram pressure. The CRe cool down by emitting radio waves via synchrotron radiation in the stripped ISM magnetic field and Inverse Comption losses with the CMB until the stripped clouds evaporate in the ICM. This way, RPS contributes to the accumulation of low-energy CRe and magnetic fields in the ICM which are available as seed-electrons for re-acceleration by ICM turbulence and shocks \citep{deGasperin01.2026.SKA}. The stripped tail magnetic field can be further amplified by the ICM magnetic draping \citep[][]{Dursi_2008,Pfrommer_2010, Ruszkowski_2014,Muller_2021}. In this framework, the radio tail length would be mainly driven by two factors, the CRe cooling time \citep[][]{Pacholczyk_1970}, and the radio plasma bulk velocity along the stripping direction. The synchrotron emission frequency, $\nu$, of a single CRe depends on its energy, $E$, and the magnetic field intensity, $B$, as $\nu\propto E^2B$. Therefore, under the assumption that the relativistic electrons move in a uniform velocity bulk motion in a uniform magnetic field, then the tail scale length $D$ would decrease with the observed frequency as $D\propto\nu^{-1/2}$ \citep[][]{Ignesti_2021}. 

This basic relation entails that to grasp the full tail extent is necessary to observe them at low frequencies. An interesting implication of this framework is that the flux density gradient along the tail, for a given magnetic field intensity, depends on the stripped ISM velocity. Therefore, by fitting the spectral curvature across wideband data, it is possible to constrain the stripped ISM velocity along the plane of the sky and, correspondingly, the galaxy velocity with respect to the ICM \citep[][]{Ignesti_2023, Roberts_2024a,Roberts_2024}, which makes radio continuum observations a valuable tool to constrain the galaxy dynamics in the cluster. Finally, LOFAR has proven that radio continuum tails are observed also in galaxies without clear evidence of ongoing RPS in the optical band, which instead appear only in the most extreme cases. Moreover, the large primary beam at low frequency allows us to map the entire cluster volume, which further contributes in making radio-selected samples less biased than more targeted spectroscopic observations. Therefore, radio-selected RPS galaxies can be adopted to build less biased samples for follow-up optical studies.

A current open question regards the properties of the magnetic field and its role in regulating the ISM-ICM mixing. The stripped ISM, with typical temperature of $10^{2-5}$ K, \citep[][]{Spitzer_1978} can interact with the ICM, which is a weakly magnetized plasma characterized by a density of $10^{-4}-10^{-3}$ particles cm$^{-3}$, a temperature of $10^{7-8}$ K \citep[][]{Sarazin1988}, and $\mu$G-level magnetic fields \citep[][]{Govoni2004}. The large temperature and velocity differences between the two phases imply short evaporation timescales for the stripped ISM \citep[$\sim10^{7-8}$ yr,][]{Klein1994} due to a combination of hydrodynamical instabilities and thermal conduction. As the typical stripping timescales are an order of magnitude larger \citep[$\sim10^{8-9}$ yr,][]{Smith2022b,Rohr_2023}, the expectation is that the stripped ISM will completely evaporate in the ICM while being stripped. Yet, in jellyfish galaxies, we observe trails of stripped ISM extending for tens of kpc and hosting active star-forming regions. This result proves the existence of a mechanism that can stabilize the stripped ISM, thus extending its survival outside of the stellar disk and permitting it to cool down and form new stars. A possible explanation for the ISM cooling in the hostile ICM conditions is the presence of ordered magnetic fields at the hot-cold plasmas interface, which can dampen the thermal conduction between the hot and cold phases and stabilize against hydrodynamical instabilities \citep[][]{McCourt_2015,Sparre2020,Sparre2024,Sparre_2024}, as well as reducing the gas mass loss \citep[][]{Rintoul_2025}. 

The presence of magnetic fields in the stripped material is naturally expected due to both internal and external factors. On the one hand, the stripped material is expected to be magnetized because it contains the ISM magnetic field bound to the thermal gas being removed by the ram pressure \citep[][]{Vollmer_2004,Ignesti2023, Vollmer_2024}. In this scenario, due to the turbulent, small-scale motions in the stripped material \citep[][]{Li_2023,Ignesti_2024,Choi_2026} the stripped tail is expected to show a low degree of polarized synchrotron emission, as consequence of the magnetic field disordered structure and the strong Faraday depolarization resulting from the thermal plasma mixed with the nonthermal components. 

On the other hand, the weak magnetic field permeating the ICM can accumulate around the infalling galaxy via the so-called magnetic draping \citep[][]{Dursi2008,Pfrommer2010}, which would naturally provide a way for jellyfish galaxies to surround themselves with large-scale magnetic fields accreted from the environment. Numerical simulations \citep[][]{Dursi2008} indicate that a prerequisite for the formation of the large-scale, ordered magnetic drape is that the galaxy's velocity must exceed the local Alfv\'en speed $V_A=B/\sqrt{4\pi\rho}$, where $B$ is the magnetic field and $\rho$ is the ion mass density, to bend the external magnetic field on its surface. This condition is virtually always satisfied in galaxy clusters, where the typical ICM Alfv\'en speed is of the order of several tens of km s$^{-1}$ and the cluster velocity dispersion is typically of the order of several hundreds of km s$^{-1}$ \citep[][]{Girardi_1993}. Furthermore, in the case of supersonic motion, the ICM magnetic field can be significantly amplified at the curved bow-shock propagating into an inhomogeneous ICM, which adiabatically enhances the ICM magnetic field via shock compression and injects turbulence that could further amplify the magnetic field via a small-scale dynamo. In fact, for galaxies moving supersonically in the ICM, numerical simulations predict the formation of an ordered magnetic drape extending for tens of kpc in the galaxy wake \citep[][]{Sparre2020,Sparre2024}. This mechanism would be the one responsible for the formation of an ordered field ``shielding'' the stripped material. In this scenario, the galaxy is also expected to show a high degree of polarized emission thanks to the magnetic field being ordered on large scales and the fact that it would be less affected by the Faraday depolarization induced by the stripped material. As in galaxy clusters, the speed of sound is comparable to the cluster velocity dispersion, supersonic draping is expected to be at work for jellyfish galaxies, which typically are the fastest cluster galaxies \citep[][]{Jaffe_2018}, providing a potential explanation for the origin of the long star-forming tails \citep{Ignesti_2026}. On the contrary, due to the mostly unknown properties of galaxy groups' magnetic fields, the role of magnetic draping in less massive halos is still unexplored. However, detecting polarized emission from the RPS tails is challenging and it requires balancing two competing factors: the surrounding ICM Faraday rotation, which depolarizes the emission, especially at low frequencies, and the intrinsic steep-spectrum nature of the radio continuum tails with a scale length decreasing with the CRe energy, which prevents their detection at high frequencies. Therefore, we currently lack a census of polarized properties for RPS galaxies that could help us test the model.

\section{Studying the physics of ram pressure stripping in the SKA era}

\begin{figure}[h!]
\centering
    \includegraphics[width=.7\linewidth]{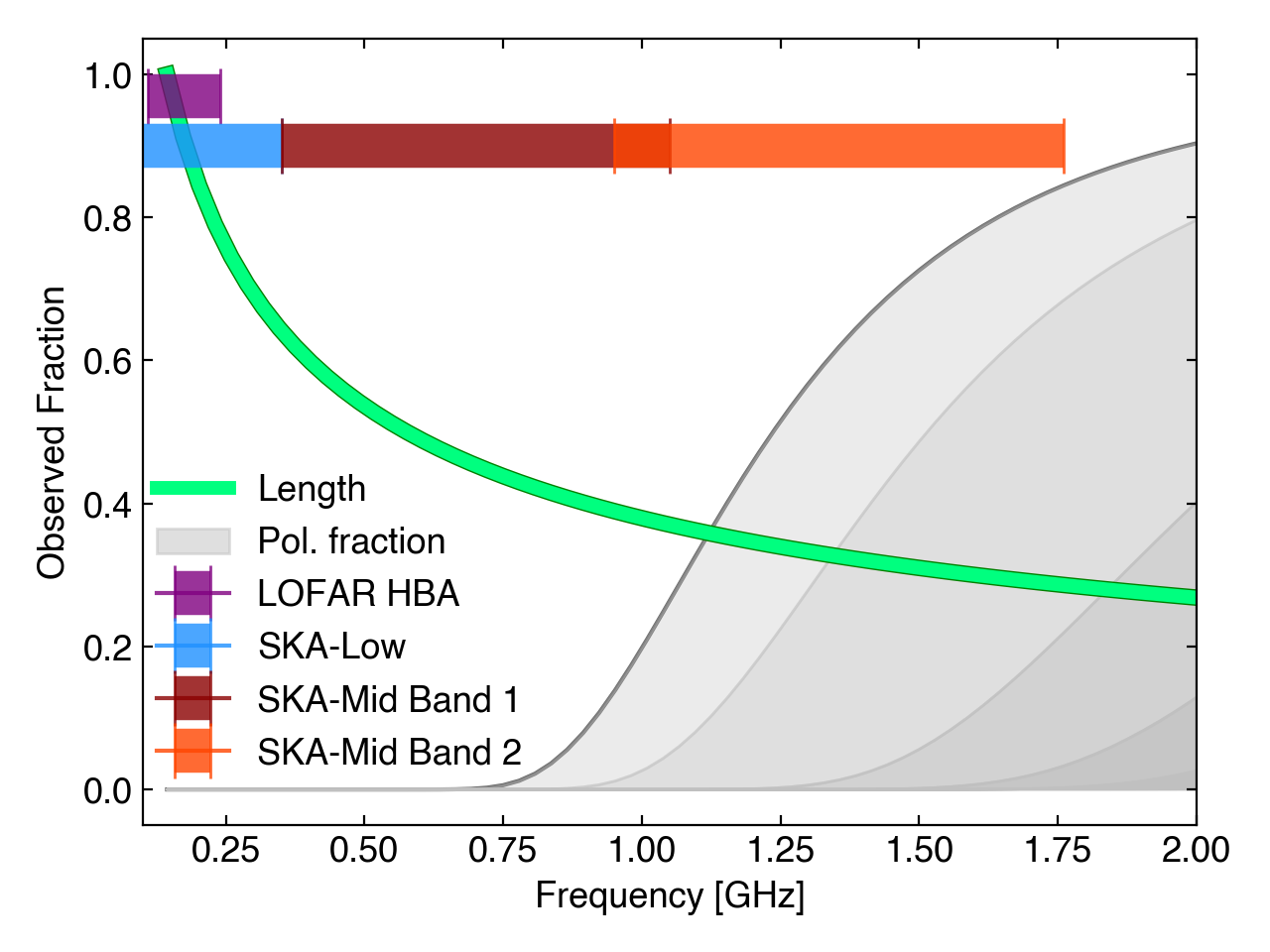}
    \includegraphics[width=0.7\linewidth]{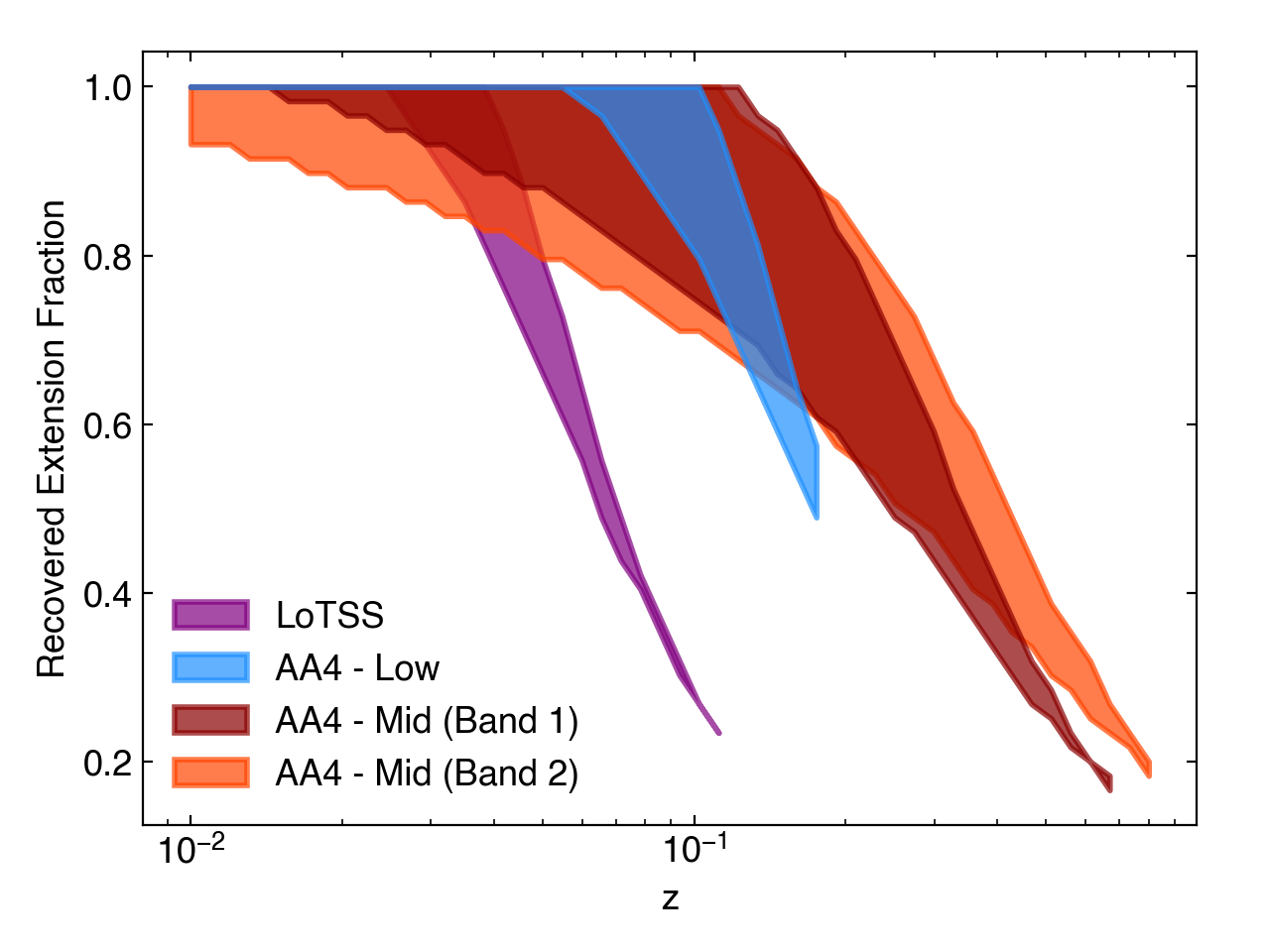}
    \caption{\textit{Top: } Reference evolution of radio tail scale length (green) and polarization fraction (silver, darker shades correspond to increasing ICM rotation measure levels of -10, -15, -30, -45,  and -60 rad m$^{-2}$) with the observed frequency. The horizontal bars indicate the frequency band observed by LOFAR HBA (purple), SKA-Low (blue), SKA-Mid band 1 (red) and band 2 (orange). \textit{Bottom: }Recovered RPS radio tail observed fraction by LOFAR LoTSS and the different SKA configurations at increasing redshift. For each configuration, we report here the angular resolution, $\theta$, and the expected RMS, $\sigma$: LoTSS $\theta=6$ arcseconds, $\sigma=100~\mu\text{Jy beam}^{-1}$; AA4-Low $\theta=9$ arcseconds, $\sigma=12~\mu\text{Jy beam}^{-1}$; AA4-MID (B1) $\theta=1$ arcseconds, $\sigma=2~\mu\text{Jy beam}^{-1}$; AA4-MID (B2) $\theta=0.8$ arcseconds, $\sigma=1~\mu\text{Jy beam}^{-1}$}
    \label{ska-test}
\end{figure}

The Square Kilometre Array will be crucial to push forward the current studies on RPS. Although providing exact estimates on its impact is not trivial, due to the transitory nature of the RPS phase for cluster galaxies, where the duration depends on the infalling orbit and the galaxy mass, and the intrinsic dependence on projection effects, which can hide the RPS features, in the following the provide a framework to evaluate the potential impact of SKA observations. The combination of the intrinsic radio tail properties and the ICM depolarization makes that the full radio tail extension can be observed mainly below 200 MHz, whereas its polarized emission becomes visible above GHz frequencies. This behavior is exemplified in Figure \ref{ska-test} (top panel), where we compare the simplified radio tail spectral contraction \citep{Ignesti_2021} with the corresponding ICM depolarization for typical ICM conditions (RM$>10$ rad m$^{-2}$) at increased observed frequency. From this comparison, it emerges that, for moderate ICM depolarization, the radio band around 1 GHz could have the optimal conditions to observe both a significant tail extent with, potentially, a non-negligible polarization fraction. Moreover, due to the high angular resolution, SKA-Mid Band 1 observations will be able to resolve the magnetic field structure on sub-kpc scale for $z<0.05$, thus future observations could be able to detect the small-scale polarized emission associated with the stripped ISM magnetic field and, possibly, disentangle it from the large-scale polarized emission associated with the external ICM drape. When taking into consideration the frequency bands that will be covered by SKA, it becomes clear that SKA-Low and, to a lower extent, SKA-Mid Band 1 will be able to observe most of their the radio tail's full extent, hence facilitating locating RPS candidates in wide-field observations. Moreover, SKA-Low will operate in the southern sky with a greater sensitivity and bandwidth than LOFAR, with a resolution comparable to the current wide-area surveys, greatly increasing the number of galaxies with evidence for ongoing RPS. Having access to a large sample of RPS galaxies will permit us, for the first time, to assess the effects of the cluster/group dynamics on RPS galaxy fraction \citep[][]{Lourenco_2023}. Finally, opening the southern sky will be important for multi-wavelength studies as well because it will permit synergies with the Southern Observatories such as VLT, ALMA and ELT, which are characterized by high angular resolution but small field of view, hence they are less suited for wide-field surveys. Combining these instruments will permit us to carry out multi-wavelength, high-resolution follow-up observations for new RPS candidates found by SKA-Low to assess how the different ISM components, nonthermal, molecular and ionized, evolve under the effect of RPS. We also note that an important feature of studying RPS in dense environments is that they greatly synergize with dedicated galaxy clusters and groups studies, such as the search for diffuse radio continuum emission on Mpc scales, or polarimetric surveys that will map the clusters' magnetic fields, representing an excellent ancillary case for future, wide-field observations which can lead to serendipitous detection of new RPS candidates.

Concerning the study of polarized emission, with the exciting prospect of finally exploring the elusive extraplanar magnetic fields, SKA-Mid Band 2 will fall in the ’sweet spot' at 1.4 GHz in which it will be possible to detect a significant fraction of both the radio tail extent and its intrinsic polarized emission. Therefore, SKA-Mid will be able to explore two crucial open questions, map the magnetic field of RPS galaxies, potentially down to sub-kpc scale thanks to the sub-arcsecond resolution achievable in AA4 configuration \citep{Loi01.2026.SKA}, and investigate the interplay between neutral and relativistic ISM components under ram pressure. The first will address our primary open questions and, regarding the latter, observation at 1.4 GHz will simultaneously trace the synchrotron nonthermal radio emission and the neutral hydrogen 21-cm radiation, which is a crucial tracer for environmental processing \citep[for a detailed discussion, we refer to][]{Ramatsoku01.2026.SKA}.

Combining SKA Low and Mid observations will allow us to routinely measure the spectral properties of RPS tails across a large bandwidth from 200 to 2000 MHz. This will allow us to constrain the dynamics of the infalling galaxies and the time scales of the RPS for large samples. For reference, an exposure time of 1 hr with SKA-Low and SKA-Mid band 2 in AA4 configuration taken in the full corresponding bandwidths\footnote{\url{https://sensitivity-calculator.skao.int/}} will result in continuum noise levels at 200 and 1300 MHz of, respectively, 11.1 and 2.7 $\mu$Jy beam$^{-1}$ for similar angular resolution of 9 arcseconds with \texttt{robust=0}, which would allow us to recover all the emission with a spectral $\alpha\geq-0.8$\footnote{The synchrotron spectral index is defined as $S(\nu)\propto\nu^\alpha$, where $S(\nu)$ is the flux density observed at the frequency $\nu$ and $\alpha$ is the spectral index.}, typically associated with the non-thermal plasma located at a few kpc from the stellar disk \citep[][]{Ignesti_2023}. The proposed SKA multi-band study, which can also probe the recent star formation rate on different time-scales \citep[][]{Condon_1992,Kennicutt_2012}, could be further combined with studies of star formation history in the optical spectrum, which provide the star formation quenching time-scale, and with HI imaging, which measures the ISM fraction removed from the galaxy to address the open question of the effectiveness and speed of quenching from RPS in dense environment.

The SKA, thanks to the unprecedented combination of resolution and sensitivity, will also permit us to explore larger cosmological volumes than is currently possible. To assess its potential in detecting RPS tails at increasing cosmological distances, in Figure \ref{ska-test} (bottom panel), we show the observable fraction of a typical RPS tail located as a function of the redshift that would be recovered in 1 hr observation by SKA. Here we propose a simplified framework in which we use the semi-empirical radio tail profile presented in \citet{Ignesti_2023} to simulate two extreme possibilities, a small tail with a starting luminosity, which is the luminosity at the stellar disk, of $10^{26}$ erg s$^{-1}$ Hz$^{-1}$ and a stripping velocity of 400 km s$^{-1}$, and an extreme case with an initial luminosity of $10^{28}$ erg s$^{-1}$ Hz$^{-1}$ and stripping velocity of 800 km s$^{-1}$. For simplicity, we assume the magnetic field for which the radiative losses are minimized $B\simeq1.9\cdot(1+z)^2$ $\mu$G and we neglect the potential role of projection effects on the observed tail projected length. The flux density profile predicted by the model is then converted into a surface brightness profile under the simplified assumption that the tail width is always resolved, and hence that the only significant dimension affected by the cosmological distance is the tail projected length. The resulting surface brightness profiles at increasing redshift $z$ are then compared with the expected sensitivity and resolution that will be achieved by SKA in AA4 configuration to determine which is the largest angular distance at which the radio tail can be considered as marginally resolved, which we define as the maximum angular distance at which the emission can be detected with a signal-to-noise ratio higher than three and for which the corresponding angular size, not counting for the stellar disk, is larger than the angular resolution. For reference, we include in this exercise also the sensitivity and resolution currently achieved by the LOFAR Two-metre Sky Survey \citep[LoTSS,][]{Shimwell_2017}, which provided us the LoTSS Jellyfish Galaxies sample \citep[][]{Roberts_2021a,Roberts_2021b}. It emerges that LOFAR has been formidable in detecting radio tails up to $z\simeq0.05$, but the limited resolution and sensitivity makes ti difficult to resolve radio tails, and hence locate RPS candidates, beyond $z=0.1$. We note, however, that deep LOFAR pointings \citep[][]{Botteon_2022} have managed to detect RPS tails up to $z=0.08$ \citep{Ignesti_2023}. In the AA4 configuration, SKA-Low will permit us to recover, at least, 50$\%$ of the radio tail lengths up to $z\simeq0.1$, thus expanding the cosmological volume previously probed by LOFAR by a factor $\sim9$, with the advantage of observing a different part of the sky as discussed above. For reference, by considering the 4955 clusters at $z<0.1$ with mass higher than $5\times10^{13}$ $M_{\odot}$, which contain a total of 67759 member galaxies \citep{Wen_2024}, based on the typical number of RPS candidates per cluster \citep[$\sim9\%$,][]{Vulcani_2023}, we can expect to detect hundreds of new candidates, effectively improving the current statistic by more than an order of magnitude. SKA-Mid, albeit suffering from the intrinsic observed tail decrease resulting from the higher observed frequency, thanks to the higher angular resolution and sensitivity, will be able to resolve the RPS radio tails up to $z\simeq0.5$. The increased cosmological volume that will be probed by SKA will further expand the RPS galaxy statistics in clusters and groups, and will potentially help us to study RPS in more rare, extreme environments such as the most massive galaxy clusters or the Cosmic Web filaments. These future high-redshift radio-selected samples will be an ideal target to be explored with a wide-field, high-angular resolution optical survey, such as the EUCLID mission \citep{EUCLID_2025}  the LSST survey at Vera C. Rubin telescope \citep[][]{LSST_2019}, and the next-generation Roman Space Telescope \citep{Roman_2025}, or with the future high-angular resolution camera on the ELT \citep[][]{Davies_2010}. Future SKA developments, such as the construction of longer baselines for SKA-Low, could further increase the importance of SKA in future studies. Reaching sub-kpc physical resolution would allow us to conduct detailed studies of the resolved nonthermal properties, such as high-resolution spectral index mapping or small-scale H$\alpha$-radio continuum comparison to evaluate how the local ISM-ICM properties or the star formation spatial distribution influence the nonthermal ISM properties.

\section{Conclusions}
The study of star-forming galaxies in clusters and groups permits us to explore both how the environment shapes the evolution of galaxies and how the ISM evolves under extreme conditions, providing us with important insights to interpret the evolution of multi-phase plasmas.  At the same time, these studies bridge between clusters and galaxies studies, radio and optical observations, being thus the crossroads of many astrophysical cases. Radio continuum studies have demonstrated to play a key role by opening us a window on the nonthermal ISM and by providing us with large, unbiased samples of RPS candidates. The SKA's crucial contribution will be in the following areas: I) greatly increase the statistics of RPS candidates both by opening the southern sky at low frequencies, thanks to SKA-Low, and by expanding the explored cosmological volume from $z\simeq0.05$ to $z\simeq0.5$, thanks to SKA-Mid. The crucial improvement in the sample size, which could be achieved by the end of the AA$^*$ phase thanks to opening of the Southern sky at low frequencies with SKA-Low, will permit us to build new samples for follow-up multi-wavelength studies; II) map the extraplanar magnetic fields with kpc-scale resolution, hence exploring the extraplanar magnetic fields elusive properties; III) routinely detect the stripped plasma synchrotron emission down to $\alpha\gtrsim-1$, providing us a deep view on the stripped nonthermal ISM evolution and its role in the context of ICM microphysics studies. These results will be transformative for the field, both in terms of an increase in statistics and in physical details of the stripped ISM evolution. An important feature of these studies is that they will greatly synergize with galaxy cluster and group studies, representing an excellent ancillary case for future observations. 

\bibliographystyle{abbrvnat-maxbibnames4}
\bibliography{chapter} 

@ARTICLE{Barnes_1992,
       author = {{Barnes}, Joshua E. and {Hernquist}, Lars},
        title = "{Dynamics of interacting galaxies.}",
      journal = {\araa},
         year = 1992,
        month = jan,
       volume = {30},
        pages = {705-742},
          doi = {10.1146/annurev.aa.30.090192.003421},
       adsurl = {https://ui.adsabs.harvard.edu/abs/1992ARA&A..30..705B}
}

@ARTICLE{EUCLID_2025,
       author = {{Euclid Collaboration} and {Mellier}, Y. and {Abdurro'uf} and {Acevedo Barroso}, J.~A. and {Ach{\'u}carro}, A. and {Adamek}, J. and {Adam}, R. and {Addison}, G.~E. and {Aghanim}, N. and {Aguena}, M. and {Ajani}, V. and {Akrami}, Y. and {Al-Bahlawan}, A. and {Alavi}, A. and {Albuquerque}, I.~S. and {Alestas}, G. and {Alguero}, G. and {Allaoui}, A. and {Allen}, S.~W. and {Allevato}, V. and {Alonso-Tetilla}, A.~V. and {Altieri}, B. and {Alvarez-Candal}, A. and {Alvi}, S. and {Amara}, A. and {Amendola}, L. and {Amiaux}, J. and {Andika}, I.~T. and {Andreon}, S. and {Andrews}, A. and {Angora}, G. and {Angulo}, R.~E. and {Annibali}, F. and {Anselmi}, A. and {Anselmi}, S. and {Arcari}, S. and {Archidiacono}, M. and {Aric{\`o}}, G. and {Arnaud}, M. and {Arnouts}, S. and {Asgari}, M. and {Asorey}, J. and {Atayde}, L. and {Atek}, H. and {Atrio-Barandela}, F. and {Aubert}, M. and {Aubourg}, E. and {Auphan}, T. and {Auricchio}, N. and {Aussel}, B. and {Aussel}, H. and {Avelino}, P.~P. and {Avgoustidis}, A. and {Avila}, S. and {Awan}, S. and {Azzollini}, R. and {Baccigalupi}, C. and {Bachelet}, E. and {Bacon}, D. and {Baes}, M. and {Bagley}, M.~B. and {Bahr-Kalus}, B. and {Balaguera-Antolinez}, A. and {Balbinot}, E. and {Balcells}, M. and {Baldi}, M. and {Baldry}, I. and {Balestra}, A. and {Ballardini}, M. and {Ballester}, O. and {Balogh}, M. and {Ba{\~n}ados}, E. and {Barbier}, R. and {Bardelli}, S. and {Baron}, M. and {Barreiro}, T. and {Barrena}, R. and {Barriere}, J.-C. and {Barros}, B.~J. and {Barthelemy}, A. and {Bartolo}, N. and {Basset}, A. and {Battaglia}, P. and {Battisti}, A.~J. and {Baugh}, C.~M. and {Baumont}, L. and {Bazzanini}, L. and {Beaulieu}, J.-P. and {Beckmann}, V. and {Belikov}, A.~N. and {Bel}, J. and {Bellagamba}, F. and {Bella}, M. and {Bellini}, E. and {Benabed}, K. and {Bender}, R. and {Benevento}, G. and {Bennett}, C.~L. and {Benson}, K. and {Bergamini}, P. and {Bermejo-Climent}, J.~R. and {Bernardeau}, F. and {Bertacca}, D. and {Berthe}, M. and {Berthier}, J. and {Bethermin}, M. and {Beutler}, F. and {Bevillon}, C. and {Bhargava}, S. and {Bhatawdekar}, R. and {Bianchi}, D. and {Bisigello}, L. and {Biviano}, A. and {Blake}, R.~P. and {Blanchard}, A. and {Blazek}, J. and {Blot}, L. and {Bosco}, A. and {Bodendorf}, C. and {Boenke}, T. and {B{\"o}hringer}, H. and {Boldrini}, P. and {Bolzonella}, M. and {Bonchi}, A. and {Bonici}, M. and {Bonino}, D. and {Bonino}, L. and {Bonvin}, C. and {Bon}, W. and {Booth}, J.~T. and {Borgani}, S. and {Borlaff}, A.~S. and {Borsato}, E. and {Bose}, B. and {Botticella}, M.~T. and {Boucaud}, A. and {Bouche}, F. and {Boucher}, J.~S. and {Boutigny}, D. and {Bouvard}, T. and {Bouwens}, R. and {Bouy}, H. and {Bowler}, R.~A.~A. and {Bozza}, V. and {Bozzo}, E. and {Branchini}, E. and {Brando}, G. and {Brau-Nogue}, S. and {Brekke}, P. and {Bremer}, M.~N. and {Brescia}, M. and {Breton}, M.-A. and {Brinchmann}, J. and {Brinckmann}, T. and {Brockley-Blatt}, C. and {Brodwin}, M. and {Brouard}, L. and {Brown}, M.~L. and {Bruton}, S. and {Bucko}, J. and {Buddelmeijer}, H. and {Buenadicha}, G. and {Buitrago}, F. and {Burger}, P. and {Burigana}, C. and {Busillo}, V. and {Busonero}, D. and {Cabanac}, R. and {Cabayol-Garcia}, L. and {Cagliari}, M.~S. and {Caillat}, A. and {Caillat}, L. and {Calabrese}, M. and {Calabro}, A. and {Calderone}, G. and {Calura}, F. and {Camacho Quevedo}, B. and {Camera}, S. and {Campos}, L. and {Ca{\~n}as-Herrera}, G. and {Candini}, G.~P. and {Cantiello}, M. and {Capobianco}, V. and {Cappellaro}, E. and {Cappelluti}, N. and {Cappi}, A. and {Caputi}, K.~I. and {Cara}, C. and {Carbone}, C. and {Cardone}, V.~F. and {Carella}, E. and {Carlberg}, R.~G. and {Carle}, M. and {Carminati}, L. and {Caro}, F. and {Carrasco}, J.~M. and {Carretero}, J. and {Carrilho}, P. and {Carron Duque}, J. and {Carry}, B.},
        title = "{Euclid: I. Overview of the Euclid mission}",
      journal = {\aap},
         year = 2025,
        month = may,
       volume = {697},
          eid = {A1},
        pages = {A1},
          doi = {10.1051/0004-6361/202450810},
archivePrefix = {arXiv},
       eprint = {2405.13491},
 primaryClass = {astro-ph.CO},
       adsurl = {https://ui.adsabs.harvard.edu/abs/2025A&A...697A...1E}
}

@INPROCEEDINGS{Davies_2010,
       author = {{Davies}, Richard and {Ageorges}, N. and {Barl}, L. and {Bedin}, L.~R. and {Bender}, R. and {Bernardi}, P. and {Chapron}, F. and {Clenet}, Y. and {Deep}, A. and {Deul}, E. and {Drost}, M. and {Eisenhauer}, F. and {Falomo}, R. and {Fiorentino}, G. and {F{\"o}rster Schreiber}, N.~M. and {Gendron}, E. and {Genzel}, R. and {Gratadour}, D. and {Greggio}, L. and {Grupp}, F. and {Held}, E. and {Herbst}, T. and {Hess}, H.-J. and {Hubert}, Z. and {Jahnke}, K. and {Kuijken}, K. and {Lutz}, D. and {Magrin}, D. and {Muschielok}, B. and {Navarro}, R. and {Noyola}, E. and {Paumard}, T. and {Piotto}, G. and {Ragazzoni}, R. and {Renzini}, A. and {Rousset}, G. and {Rix}, H.-W. and {Saglia}, R. and {Tacconi}, L. and {Thiel}, M. and {Tolstoy}, E. and {Trippe}, S. and {Tromp}, N. and {Valentijn}, E.~A. and {Verdoes Kleijn}, G. and {Wegner}, M.},
        title = "{MICADO: the E-ELT adaptive optics imaging camera}",
    booktitle = {Ground-based and Airborne Instrumentation for Astronomy III},
         year = 2010,
       editor = {{McLean}, Ian S. and {Ramsay}, Suzanne K. and {Takami}, Hideki},
       series = {Society of Photo-Optical Instrumentation Engineers (SPIE) Conference Series},
       volume = {7735},
        month = jul,
          eid = {77352A},
        pages = {77352A},
          doi = {10.1117/12.856379},
archivePrefix = {arXiv},
       eprint = {1005.5009},
 primaryClass = {astro-ph.IM},
       adsurl = {https://ui.adsabs.harvard.edu/abs/2010SPIE.7735E..2AD}
}

@ARTICLE{Lourenco_2023,
       author = {{Louren{\c{c}}o}, Ana C.~C. and {Jaff{\'e}}, Y.~L. and {Vulcani}, B. and {Biviano}, A. and {Poggianti}, B. and {Moretti}, A. and {Kelkar}, K. and {Crossett}, J.~P. and {Gitti}, M. and {Smith}, R. and {Lagan{\'a}}, T.~F. and {Gullieuszik}, M. and {Ignesti}, A. and {McGee}, S. and {Wolter}, A. and {Sonkamble}, S. and {M{\"u}ller}, A.},
        title = "{The effect of cluster dynamical state on ram-pressure stripping}",
      journal = {\mnras},
         year = 2023,
        month = dec,
       volume = {526},
       number = {4},
        pages = {4831-4847},
          doi = {10.1093/mnras/stad2972},
archivePrefix = {arXiv},
       eprint = {2309.15934},
 primaryClass = {astro-ph.GA},
       adsurl = {https://ui.adsabs.harvard.edu/abs/2023MNRAS.526.4831L}
}

@ARTICLE{Edler_2024,
       author = {{Edler}, H.~W. and {Roberts}, I.~D. and {Boselli}, A. and {de Gasperin}, F. and {Heesen}, V. and {Br{\"u}ggen}, M. and {Ignesti}, A. and {Gajovi{\'c}}, L.},
        title = "{ViCTORIA project: The LOFAR view of environmental effects in Virgo cluster star-forming galaxies}",
      journal = {\aap},
         year = 2024,
        month = mar,
       volume = {683},
          eid = {A149},
        pages = {A149},
          doi = {10.1051/0004-6361/202348301},
archivePrefix = {arXiv},
       eprint = {2311.01904},
 primaryClass = {astro-ph.GA},
       adsurl = {https://ui.adsabs.harvard.edu/abs/2024A&A...683A.149E}
}

@ARTICLE{Lal_2022,
       author = {{Lal}, D.~V. and {Lyskova}, N. and {Zhang}, C. and {Venturi}, T. and {Forman}, W.~R. and {Jones}, C. and {Churazov}, E.~M. and {van Weeren}, R.~J. and {Bonafede}, A. and {Miller}, N.~A. and {Roberts}, I.~D. and {Bykov}, A.~M. and {Di Mascolo}, L. and {Br{\"u}ggen}, M. and {Brunetti}, G.},
        title = "{High-resolution, High-sensitivity, Low-frequency uGMRT View of Coma Cluster of Galaxies}",
      journal = {\apj},
         year = 2022,
        month = aug,
       volume = {934},
       number = {2},
          eid = {170},
        pages = {170},
          doi = {10.3847/1538-4357/ac7a9b},
archivePrefix = {arXiv},
       eprint = {2207.06624},
 primaryClass = {astro-ph.GA},
       adsurl = {https://ui.adsabs.harvard.edu/abs/2022ApJ...934..170L}
}

@ARTICLE{Beck_2005,
       author = {{Beck}, R. and {Krause}, M.},
        title = "{Revised equipartition and minimum energy formula for magnetic field strength estimates from radio synchrotron observations}",
      journal = {Astronomische Nachrichten},
         year = 2005,
        month = jul,
       volume = {326},
       number = {6},
        pages = {414-427},
          doi = {10.1002/asna.200510366},
archivePrefix = {arXiv},
       eprint = {astro-ph/0507367},
 primaryClass = {astro-ph},
       adsurl = {https://ui.adsabs.harvard.edu/abs/2005AN....326..414B}
}

@ARTICLE{Franchetto_2021,
       author = {{Franchetto}, Andrea and {Tonnesen}, Stephanie and {Poggianti}, Bianca M. and {Vulcani}, Benedetta and {Gullieuszik}, Marco and {Moretti}, Alessia and {Smith}, Rory and {Ignesti}, Alessandro and {Bacchini}, Cecilia and {McGee}, Sean and {Tomi{\v{c}}i{\'c}}, Neven and {Mingozzi}, Matilde and {Wolter}, Anna and {M{\"u}ller}, Ancla},
        title = "{Evidence for Mixing between ICM and Stripped ISM by the Analysis of the Gas Metallicity in the Tails of Jellyfish Galaxies}",
      journal = {\apjl},
         year = 2021,
        month = nov,
       volume = {922},
       number = {1},
          eid = {L6},
        pages = {L6},
          doi = {10.3847/2041-8213/ac3664},
archivePrefix = {arXiv},
       eprint = {2111.04755},
 primaryClass = {astro-ph.GA},
       adsurl = {https://ui.adsabs.harvard.edu/abs/2021ApJ...922L...6F}
}

@incollection{deGasperin01.2026.SKA, author = {Francesco de Gasperin and author2 and author3 and author4 and author5},title = {},year = {2026},publisher = {},note = {arXiv search: Report number AASKAII/deGasperin01},booktitle = {Advancing Astrophysics with the SKA -- II (AASKAII)}}

@incollection{Ramatsoku01.2026.SKA, author = {Mpati Ramatsoku and author2 and author3 and author4 and author5},title = {},year = {2026},publisher = {},note = {arXiv search: Report number AASKAII/Ramatsoku01},booktitle = {Advancing Astrophysics with the SKA -- II (AASKAII)}}

@incollection{Loi01.2026.SKA, author = {Francesca Loi and author2 and author3 and author4 and author5},title = {},year = {2026},publisher = {},note = {arXiv search: Report number AASKAII/Loi01},booktitle = {Advancing Astrophysics with the SKA -- II (AASKAII)}}

@ARTICLE{Choi_2026,
       author = {{Choi}, Woorak and {Chung}, Aeree and {Kim}, Chang-Goo and {Lee}, Bumhyun and {Cortese}, Luca and {Brown}, Toby and {Catinella}, Barbara and {Emsellem}, Eric and {Fraser-McKelvie}, A. and {Sun}, Jiayi and {Watts}, Adam},
        title = "{The Impact of Ram Pressure on the Radio Spectral Index and Magnetic Field of NGC 4522: A High-resolution VLA Continuum Study}",
      journal = {\apj},
         year = 2026,
        month = mar,
       volume = {999},
       number = {1},
          eid = {116},
        pages = {116},
          doi = {10.3847/1538-4357/ae39ca},
archivePrefix = {arXiv},
       eprint = {2601.10803},
 primaryClass = {astro-ph.GA},
       adsurl = {https://ui.adsabs.harvard.edu/abs/2026ApJ...999..116C}
}

@ARTICLE{Li_2023,
       author = {{Li}, Yuan and {Luo}, Rongxin and {Fossati}, Matteo and {Sun}, Ming and {J{\'a}chym}, Pavel},
        title = "{Turbulence in the tail of a jellyfish galaxy}",
      journal = {\mnras},
         year = 2023,
        month = may,
       volume = {521},
       number = {3},
        pages = {4785-4791},
          doi = {10.1093/mnras/stad874},
archivePrefix = {arXiv},
       eprint = {2303.15500},
 primaryClass = {astro-ph.GA},
       adsurl = {https://ui.adsabs.harvard.edu/abs/2023MNRAS.521.4785L}
}

@ARTICLE{LSST_2019,
       author = {{Ivezi{\'c}}, {\v{Z}}eljko and {Kahn}, Steven M. and {Tyson}, J. Anthony and {Abel}, Bob and {Acosta}, Emily and {Allsman}, Robyn and {Alonso}, David and {AlSayyad}, Yusra and {Anderson}, Scott F. and {Andrew}, John and {Angel}, James Roger P. and {Angeli}, George Z. and {Ansari}, Reza and {Antilogus}, Pierre and {Araujo}, Constanza and {Armstrong}, Robert and {Arndt}, Kirk T. and {Astier}, Pierre and {Aubourg}, {\'E}ric and {Auza}, Nicole and {Axelrod}, Tim S. and {Bard}, Deborah J. and {Barr}, Jeff D. and {Barrau}, Aurelian and {Bartlett}, James G. and {Bauer}, Amanda E. and {Bauman}, Brian J. and {Baumont}, Sylvain and {Bechtol}, Ellen and {Bechtol}, Keith and {Becker}, Andrew C. and {Becla}, Jacek and {Beldica}, Cristina and {Bellavia}, Steve and {Bianco}, Federica B. and {Biswas}, Rahul and {Blanc}, Guillaume and {Blazek}, Jonathan and {Blandford}, Roger D. and {Bloom}, Josh S. and {Bogart}, Joanne and {Bond}, Tim W. and {Booth}, Michael T. and {Borgland}, Anders W. and {Borne}, Kirk and {Bosch}, James F. and {Boutigny}, Dominique and {Brackett}, Craig A. and {Bradshaw}, Andrew and {Brandt}, William Nielsen and {Brown}, Michael E. and {Bullock}, James S. and {Burchat}, Patricia and {Burke}, David L. and {Cagnoli}, Gianpietro and {Calabrese}, Daniel and {Callahan}, Shawn and {Callen}, Alice L. and {Carlin}, Jeffrey L. and {Carlson}, Erin L. and {Chandrasekharan}, Srinivasan and {Charles-Emerson}, Glenaver and {Chesley}, Steve and {Cheu}, Elliott C. and {Chiang}, Hsin-Fang and {Chiang}, James and {Chirino}, Carol and {Chow}, Derek and {Ciardi}, David R. and {Claver}, Charles F. and {Cohen-Tanugi}, Johann and {Cockrum}, Joseph J. and {Coles}, Rebecca and {Connolly}, Andrew J. and {Cook}, Kem H. and {Cooray}, Asantha and {Covey}, Kevin R. and {Cribbs}, Chris and {Cui}, Wei and {Cutri}, Roc and {Daly}, Philip N. and {Daniel}, Scott F. and {Daruich}, Felipe and {Daubard}, Guillaume and {Daues}, Greg and {Dawson}, William and {Delgado}, Francisco and {Dellapenna}, Alfred and {de Peyster}, Robert and {de Val-Borro}, Miguel and {Digel}, Seth W. and {Doherty}, Peter and {Dubois}, Richard and {Dubois-Felsmann}, Gregory P. and {Durech}, Josef and {Economou}, Frossie and {Eifler}, Tim and {Eracleous}, Michael and {Emmons}, Benjamin L. and {Fausti Neto}, Angelo and {Ferguson}, Henry and {Figueroa}, Enrique and {Fisher-Levine}, Merlin and {Focke}, Warren and {Foss}, Michael D. and {Frank}, James and {Freemon}, Michael D. and {Gangler}, Emmanuel and {Gawiser}, Eric and {Geary}, John C. and {Gee}, Perry and {Geha}, Marla and {Gessner}, Charles J.~B. and {Gibson}, Robert R. and {Gilmore}, D. Kirk and {Glanzman}, Thomas and {Glick}, William and {Goldina}, Tatiana and {Goldstein}, Daniel A. and {Goodenow}, Iain and {Graham}, Melissa L. and {Gressler}, William J. and {Gris}, Philippe and {Guy}, Leanne P. and {Guyonnet}, Augustin and {Haller}, Gunther and {Harris}, Ron and {Hascall}, Patrick A. and {Haupt}, Justine and {Hernandez}, Fabio and {Herrmann}, Sven and {Hileman}, Edward and {Hoblitt}, Joshua and {Hodgson}, John A. and {Hogan}, Craig and {Howard}, James D. and {Huang}, Dajun and {Huffer}, Michael E. and {Ingraham}, Patrick and {Innes}, Walter R. and {Jacoby}, Suzanne H. and {Jain}, Bhuvnesh and {Jammes}, Fabrice and {Jee}, M. James and {Jenness}, Tim and {Jernigan}, Garrett and {Jevremovi{\'c}}, Darko and {Johns}, Kenneth and {Johnson}, Anthony S. and {Johnson}, Margaret W.~G. and {Jones}, R. Lynne and {Juramy-Gilles}, Claire and {Juri{\'c}}, Mario and {Kalirai}, Jason S. and {Kallivayalil}, Nitya J. and {Kalmbach}, Bryce and {Kantor}, Jeffrey P. and {Karst}, Pierre and {Kasliwal}, Mansi M. and {Kelly}, Heather and {Kessler}, Richard and {Kinnison}, Veronica and {Kirkby}, David and {Knox}, Lloyd and {Kotov}, Ivan V. and {Krabbendam}, Victor L. and {Krughoff}, K. Simon and {Kub{\'a}nek}, Petr and {Kuczewski}, John and {Kulkarni}, Shri and {Ku}, John and {Kurita}, Nadine R. and {Lage}, Craig S. and {Lambert}, Ron and {Lange}, Travis and {Langton}, J. Brian and {Le Guillou}, Laurent and {Levine}, Deborah and {Liang}, Ming and {Lim}, Kian-Tat and {Lintott}, Chris J. and {Long}, Kevin E. and {Lopez}, Margaux and {Lotz}, Paul J. and {Lupton}, Robert H. and {Lust}, Nate B. and {MacArthur}, Lauren A. and {Mahabal}, Ashish and {Mandelbaum}, Rachel and {Markiewicz}, Thomas W. and {Marsh}, Darren S. and {Marshall}, Philip J. and {Marshall}, Stuart and {May}, Morgan and {McKercher}, Robert and {McQueen}, Michelle and {Meyers}, Joshua and {Migliore}, Myriam and {Miller}, Michelle and {Mills}, David J.},
        title = "{LSST: From Science Drivers to Reference Design and Anticipated Data Products}",
      journal = {\apj},
         year = 2019,
        month = mar,
       volume = {873},
       number = {2},
          eid = {111},
        pages = {111},
          doi = {10.3847/1538-4357/ab042c},
archivePrefix = {arXiv},
       eprint = {0805.2366},
 primaryClass = {astro-ph},
       adsurl = {https://ui.adsabs.harvard.edu/abs/2019ApJ...873..111I}
}

@ARTICLE{Watts_2023,
       author = {{Watts}, Adam B. and {Cortese}, Luca and {Catinella}, Barbara and {Brown}, Toby and {Wilson}, Christine D. and {Zabel}, Nikki and {Roberts}, Ian D. and {Davis}, Timothy A. and {Thorp}, Mallory and {Chung}, Aeree and {Stevens}, Adam R.~H. and {Ellison}, Sara L. and {Spekkens}, Kristine and {Parker}, Laura C. and {Bah{\'e}}, Yannick M. and {Villanueva}, Vicente and {Jim{\'e}nez-Donaire}, Mar{\'\i}a and {Bisaria}, Dhruv and {Boselli}, Alessandro and {Bolatto}, Alberto D. and {Lee}, Bumhyun},
        title = "{VERTICO V: The environmentally driven evolution of the inner cold gas discs of Virgo cluster galaxies}",
      journal = {\pasa},
         year = 2023,
        month = apr,
       volume = {40},
          eid = {e017},
        pages = {e017},
          doi = {10.1017/pasa.2023.14},
archivePrefix = {arXiv},
       eprint = {2303.07549},
 primaryClass = {astro-ph.GA},
       adsurl = {https://ui.adsabs.harvard.edu/abs/2023PASA...40...17W}
}

@article{McKee-Cowie_1977,
	adsurl = "http://adsabs.harvard.edu/abs/1977ApJ...215..213M",
	author = "{McKee}, C.~F. and {Cowie}, L.~L.",
	doi = "10.1086/155350",
	journal = "\apj",
	month = jul,
	pages = "213--225",
	title = "{The evaporation of spherical clouds in a hot gas. II - Effects of radiation}",
	volume = 215,
	year = 1977
}

@ARTICLE{Smith_2022,
       author = {{Smith}, Rory and {Shinn}, Jong-Ho and {Tonnesen}, Stephanie and {Calder{\'o}n-Castillo}, Paula and {Crossett}, Jacob and {Jaffe}, Yara L. and {Roberts}, Ian and {McGee}, Sean and {George}, Koshy and {Vulcani}, Benedetta and {Gullieuszik}, Marco and {Moretti}, Alessia and {Poggianti}, Bianca M. and {Shin}, Jihye},
        title = "{A New Method to Constrain the Appearance and Disappearance of Observed Jellyfish Galaxy Tails}",
      journal = {\apj},
         year = 2022,
        month = jul,
       volume = {934},
       number = {1},
          eid = {86},
        pages = {86},
          doi = {10.3847/1538-4357/ac7ab5},
archivePrefix = {arXiv},
       eprint = {2207.00029},
 primaryClass = {astro-ph.GA},
       adsurl = {https://ui.adsabs.harvard.edu/abs/2022ApJ...934...86S}
}

@book{Pacholczyk_1970,
	adsurl = "http://adsabs.harvard.edu/abs/1970ranp.book.....P",
	author = "{Pacholczyk}, A.~G.",
	booktitle = "{Series of Books in Astronomy and Astrophysics, San Francisco: Freeman, 1970}",
	editor = "{Pacholczyk, A.~G.}",
	title = "{Radio astrophysics. Nonthermal processes in galactic and extragalactic sources}",
	year = 1970
}

@book{Spitzer_1978,
	adsurl = "http://adsabs.harvard.edu/abs/1988xrec.book.....S",
	author = "{Spitzer}, L.~Jr.",
	booktitle = "{New York: Wiley}",
	editor = "{Spitzer, L.~Jr.}",
	title = "{Physical Processes in the Interstellar Medium, 1978}",
	year = 1978
}

@ARTICLE{Vollmer_2013,
       author = {{Vollmer}, B. and {Soida}, M. and {Beck}, R. and {Chung}, A. and {Urbanik}, M. and {Chy{\.z}y}, K.~T. and {Otmianowska-Mazur}, K. and {Kenney}, J.~D.~P.},
        title = "{Large-scale radio continuum properties of 19 Virgo cluster galaxies. The influence of tidal interactions, ram pressure stripping, and accreting gas envelopes}",
      journal = {\aap},
         year = 2013,
        month = may,
       volume = {553},
          eid = {A116},
        pages = {A116},
          doi = {10.1051/0004-6361/201321163},
archivePrefix = {arXiv},
       eprint = {1304.1279},
 primaryClass = {astro-ph.CO},
       adsurl = {https://ui.adsabs.harvard.edu/abs/2013A&A...553A.116V}
}

@ARTICLE{Chen_2020,
       author = {{Chen}, Hao and {Sun}, Ming and {Yagi}, Masafumi and {Bravo-Alfaro}, Hector and {Brinks}, Elias and {Kenney}, Jeffrey and {Combes}, Francoise and {Sivanandam}, Suresh and {Jachym}, Pavel and {Fossati}, Matteo and {Gavazzi}, Giuseppe and {Boselli}, Alessandro and {Nulsen}, Paul and {Sarazin}, Craig and {Ge}, Chong and {Yoshida}, Michitoshi and {Roediger}, Elke},
        title = "{The ram pressure stripped radio tails of galaxies in the Coma cluster}",
      journal = {\mnras},
         year = 2020,
        month = aug,
       volume = {496},
       number = {4},
        pages = {4654-4673},
          doi = {10.1093/mnras/staa1868},
archivePrefix = {arXiv},
       eprint = {2004.06743},
 primaryClass = {astro-ph.GA},
       adsurl = {https://ui.adsabs.harvard.edu/abs/2020MNRAS.496.4654C}
}

@ARTICLE{Vollmer_2004,
       author = {{Vollmer}, B. and {Beck}, R. and {Kenney}, Jeffrey D.~P. and {van Gorkom}, J.~H.},
        title = "{Radio Continuum Observations of the Virgo Cluster Spiral NGC 4522: The Signature of Ram Pressure}",
      journal = {\aj},
         year = 2004,
        month = jun,
       volume = {127},
       number = {6},
        pages = {3375-3381},
          doi = {10.1086/420802},
archivePrefix = {arXiv},
       eprint = {astro-ph/0403054},
 primaryClass = {astro-ph},
       adsurl = {https://ui.adsabs.harvard.edu/abs/2004AJ....127.3375V}
}

@ARTICLE{Moretti_2020,
       author = {{Moretti}, Alessia and {Paladino}, Rosita and {Poggianti}, Bianca M. and {Serra}, Paolo and {Roediger}, Elke and {Gullieuszik}, Marco and {Tomi{\v{c}}i{\'c}}, Neven and {Radovich}, Mario and {Vulcani}, Benedetta and {Jaff{\'e}}, Yara L. and {Fritz}, Jacopo and {Bettoni}, Daniela and {Ramatsoku}, Mpati and {Wolter}, Anna},
        title = "{GASP. XXII. The Molecular Gas Content of the JW100 Jellyfish Galaxy at z {\ensuremath{\sim}} 0.05: Does Ram Pressure Promote Molecular Gas Formation?}",
      journal = {\apj},
         year = 2020,
        month = jan,
       volume = {889},
       number = {1},
          eid = {9},
        pages = {9},
          doi = {10.3847/1538-4357/ab616a},
archivePrefix = {arXiv},
       eprint = {1912.06565},
 primaryClass = {astro-ph.GA},
       adsurl = {https://ui.adsabs.harvard.edu/abs/2020ApJ...889....9M}
}

@ARTICLE{Roberts_2021a,
       author = {{Roberts}, I.~D. and {van Weeren}, R.~J. and {McGee}, S.~L. and {Botteon}, A. and {Drabent}, A. and {Ignesti}, A. and {Rottgering}, H.~J.~A. and {Shimwell}, T.~W. and {Tasse}, C.},
        title = "{LoTSS jellyfish galaxies. I. Radio tails in low redshift clusters}",
      journal = {\aap},
         year = 2021,
        month = jun,
       volume = {650},
          eid = {A111},
        pages = {A111},
          doi = {10.1051/0004-6361/202140784},
archivePrefix = {arXiv},
       eprint = {2104.05383},
 primaryClass = {astro-ph.GA},
       adsurl = {https://ui.adsabs.harvard.edu/abs/2021A&A...650A.111R}
}

@ARTICLE{Roberts_2021b,
       author = {{Roberts}, I.~D. and {van Weeren}, R.~J. and {McGee}, S.~L. and {Botteon}, A. and {Ignesti}, A. and {Rottgering}, H.~J.~A.},
        title = "{LoTSS jellyfish galaxies. II. Ram pressure stripping in groups versus clusters}",
      journal = {\aap},
         year = 2021,
        month = aug,
       volume = {652},
          eid = {A153},
        pages = {A153},
          doi = {10.1051/0004-6361/202141118},
archivePrefix = {arXiv},
       eprint = {2106.06315},
 primaryClass = {astro-ph.GA},
       adsurl = {https://ui.adsabs.harvard.edu/abs/2021A&A...652A.153R}
}

@article{Shimwell_2017,
	adsurl = "https://ui.adsabs.harvard.edu/\#abs/2017A\&A...598A.104S",
	author = "{Shimwell}, T.~W. and {R{\"o}ttgering}, H.~J.~A. and {Best}, P.~N. and {Williams}, W.~L. and {Dijkema}, T.~J. and {de Gasperin}, F. and {Hardcastle}, M.~J. and {Heald}, G.~H. and {Hoang}, D.~N. and {Horneffer}, A. and {Intema}, H. and {Mahony}, E.~K. and {Mandal}, S. and {Mechev}, A.~P. and {Morabito}, L. and {Oonk}, J.~B.~R. and {Rafferty}, D. and {Retana-Montenegro}, E. and {Sabater}, J. and {Tasse}, C. and {van Weeren}, R.~J. and {Br{\"u}ggen}, M. and {Brunetti}, G. and {Chy{\.z}y}, K.~T. and {Conway}, J.~E. and {Haverkorn}, M. and {Jackson}, N. and {Jarvis}, M.~J. and {McKean}, J.~P. and {Miley}, G.~K. and {Morganti}, R. and {White}, G.~J. and {Wise}, M.~W. and {van Bemmel}, I.~M. and {Beck}, R. and {Brienza}, M. and {Bonafede}, A. and {Calistro Rivera}, G. and {Cassano}, R. and {Clarke}, A.~O. and {Cseh}, D. and {Deller}, A. and {Drabent}, A. and {van Driel}, W. and {Engels}, D. and {Falcke}, H. and {Ferrari}, C. and {Fr{\"o}hlich}, S. and {Garrett}, M.~A. and {Harwood}, J.~J. and {Heesen}, V. and {Hoeft}, M. and {Horellou}, C. and {Israel}, F.~P. and {Kapi{\'n}ska}, A.~D. and {Kunert- Bajraszewska}, M. and {McKay}, D.~J. and {Mohan}, N.~R. and {Orr{\'u}}, E. and {Pizzo}, R.~F. and {Prandoni}, I. and {Schwarz}, D.~J. and {Shulevski}, A. and {Sipior}, M. and {Smith}, D.~J.~B. and {Sridhar}, S.~S. and {Steinmetz}, M. and {Stroe}, A. and {Varenius}, E. and {van der Werf}, P.~P. and {Zensus}, J.~A. and {Zwart}, J.~T.~L.",
	doi = "10.1051/0004-6361/201629313",
	eid = "A104",
	journal = "\aap",
	month = feb,
	pages = "A104",
	primaryclass = "astro-ph.IM",
	title = "{The LOFAR Two-metre Sky Survey. I. Survey description and preliminary data release}",
	volume = "598",
	year = 2017
}

@ARTICLE{Poggianti_2019,
       author = {{Poggianti}, Bianca M. and {Ignesti}, Alessandro and {Gitti}, Myriam and
         {Wolter}, Anna and {Brighenti}, Fabrizio and {Biviano}, Andrea and
         {George}, Koshy and {Vulcani}, Benedetta and {Gullieuszik}, Marco and
         {Moretti}, Alessia and {Paladino}, Rosita and {Bettoni}, Daniela and
         {Franchetto}, Andrea and {Jaff{\'e}}, Yara L. and {Radovich}, Mario and
         {Roediger}, Elke and {Tomi{\v{c}}i{\'c}}, Neven and
         {Tonnesen}, Stephanie and {Bellhouse}, Callum and {Fritz}, Jacopo and
         {Omizzolo}, Alessandro},
        title = "{GASP XXIII: A Jellyfish Galaxy as an Astrophysical Laboratory of the Baryonic Cycle}",
      journal = {\apj},
         year = 2019,
        month = dec,
       volume = {887},
       number = {2},
          eid = {155},
        pages = {155},
          doi = {10.3847/1538-4357/ab5224},
archivePrefix = {arXiv},
       eprint = {1910.11622},
 primaryClass = {astro-ph.GA},
       adsurl = {https://ui.adsabs.harvard.edu/abs/2019ApJ...887..155P}
}

@ARTICLE{Shimwell_2019,
       author = {{Shimwell}, T.~W. and {Tasse}, C. and {Hardcastle}, M.~J. and
         {Mechev}, A.~P. and {Williams}, W.~L. and {Best}, P.~N. and
         {R{\"o}ttgering}, H.~J.~A. and {Callingham}, J.~R. and
         {Dijkema}, T.~J. and {de Gasperin}, F. and {Hoang}, D.~N. and
         {Hugo}, B. and {Mirmont}, M. and {Oonk}, J.~B.~R. and {Prandoni}, I. and
         {Rafferty}, D. and {Sabater}, J. and {Smirnov}, O. and
         {van Weeren}, R.~J. and {White}, G.~J. and {Atemkeng}, M. and
         {Bester}, L. and {Bonnassieux}, E. and {Br{\"u}ggen}, M. and
         {Brunetti}, G. and {Chy{\.z}y}, K.~T. and {Cochrane}, R. and
         {Conway}, J.~E. and {Croston}, J.~H. and {Danezi}, A. and {Duncan}, K. and
         {Haverkorn}, M. and {Heald}, G.~H. and {Iacobelli}, M. and
         {Intema}, H.~T. and {Jackson}, N. and {Jamrozy}, M. and
         {Jarvis}, M.~J. and {Lakhoo}, R. and {Mevius}, M. and {Miley}, G.~K. and
         {Morabito}, L. and {Morganti}, R. and {Nisbet}, D. and {Orr{\'u}}, E. and
         {Perkins}, S. and {Pizzo}, R.~F. and {Schrijvers}, C. and
         {Smith}, D.~J.~B. and {Vermeulen}, R. and {Wise}, M.~W. and
         {Alegre}, L. and {Bacon}, D.~J. and {van Bemmel}, I.~M. and
         {Beswick}, R.~J. and {Bonafede}, A. and {Botteon}, A. and {Bourke}, S. and
         {Brienza}, M. and {Calistro Rivera}, G. and {Cassano}, R. and
         {Clarke}, A.~O. and {Conselice}, C.~J. and {Dettmar}, R.~J. and
         {Drabent}, A. and {Dumba}, C. and {Emig}, K.~L. and
         {En{\ss}lin}, T.~A. and {Ferrari}, C. and {Garrett}, M.~A. and
         {G{\'e}nova-Santos}, R.~T. and {Goyal}, A. and {G{\"u}rkan}, G. and
         {Hale}, C. and {Harwood}, J.~J. and {Heesen}, V. and {Hoeft}, M. and
         {Horellou}, C. and {Jackson}, C. and {Kokotanekov}, G. and
         {Kondapally}, R. and {Kunert-Bajraszewska}, M. and {Mahatma}, V. and
         {Mahony}, E.~K. and {Mandal}, S. and {McKean}, J.~P. and {Merloni}, A. and
         {Mingo}, B. and {Miskolczi}, A. and {Mooney}, S. and
         {Nikiel-Wroczy{\'n}ski}, B. and {O'Sullivan}, S.~P. and {Quinn}, J. and
         {Reich}, W. and {Roskowi{\'n}ski}, C. and {Rowlinson}, A. and
         {Savini}, F. and {Saxena}, A. and {Schwarz}, D.~J. and {Shulevski}, A. and
         {Sridhar}, S.~S. and {Stacey}, H.~R. and {Urquhart}, S. and
         {van der Wiel}, M.~H.~D. and {Varenius}, E. and {Webster}, B. and
         {Wilber}, A.},
        title = "{The LOFAR Two-metre Sky Survey. II. First data release}",
      journal = {\aap},
         year = "2019",
        month = "Feb",
       volume = {622},
          eid = {A1},
        pages = {A1},
          doi = {10.1051/0004-6361/201833559},
archivePrefix = {arXiv},
       eprint = {1811.07926},
 primaryClass = {astro-ph.GA},
       adsurl = {https://ui.adsabs.harvard.edu/abs/2019A&A...622A...1S}
}

@ARTICLE{Poggianti_2016,
       author = {{Poggianti}, B.~M. and {Fasano}, G. and {Omizzolo}, A. and
         {Gullieuszik}, M. and {Bettoni}, D. and {Moretti}, A. and
         {Paccagnella}, A. and {Jaff{\'e}}, Y.~L. and {Vulcani}, B. and
         {Fritz}, J. and {Couch}, W. and {D'Onofrio}, M.},
        title = "{Jellyfish Galaxy Candidates at Low Redshift}",
      journal = {\aj},
         year = 2016,
        month = mar,
       volume = {151},
       number = {3},
          eid = {78},
        pages = {78},
          doi = {10.3847/0004-6256/151/3/78},
archivePrefix = {arXiv},
       eprint = {1504.07105},
 primaryClass = {astro-ph.GA},
       adsurl = {https://ui.adsabs.harvard.edu/abs/2016AJ....151...78P}
}

@article{Shimwell_2022,
	title = {The {LOFAR} {Two}-metre {Sky} {Survey} ({LoTSS}). {V}. {Second} data release},
	issn = {0004-6361, 1432-0746},
	url = {https://www.aanda.org/10.1051/0004-6361/202142484},
	doi = {10.1051/0004-6361/202142484},
	urldate = {2022-01-31},
	author = {Shimwell, T. W. and Hardcastle, M. J. and Tasse, C. and Best, P. N. and Röttgering, H. J. A. and Williams, W. L. and Botteon, A. and Drabent, A. and Mechev, A. and Shulevski, A. and van Weeren, R. J. and {al.}},
	journal = {\aap},
	month = jan,
	year = {2022},
}

@ARTICLE{Roberts_2021c,
       author = {{Roberts}, I.~D. and {van Weeren}, R.~J. and {Timmerman}, R. and {Botteon}, A. and {Gendron-Marsolais}, M. and {Ignesti}, A. and {Rottgering}, H.~J.~A.},
        title = "{LoTSS jellyfish galaxies. III. The first identification of jellyfish galaxies in the Perseus cluster}",
      journal = {\aap},
         year = 2022,
        month = feb,
       volume = {658},
          eid = {A44},
        pages = {A44},
          doi = {10.1051/0004-6361/202142294},
archivePrefix = {arXiv},
       eprint = {2112.08728},
 primaryClass = {astro-ph.GA},
       adsurl = {https://ui.adsabs.harvard.edu/abs/2022A&A...658A..44R}
}

@ARTICLE{Peluso_2021,
       author = {{Peluso}, Giorgia and {Vulcani}, Benedetta and {Poggianti}, Bianca M. and {Moretti}, Alessia and {Radovich}, Mario and {Smith}, Rory and {Jaff{\'e}}, Yara L. and {Crossett}, Jacob and {Gullieuszik}, Marco and {Fritz}, Jacopo and {Ignesti}, Alessandro},
        title = "{Exploring the AGN-Ram Pressure Stripping Connection in Local Clusters}",
      journal = {\apj},
         year = 2022,
        month = mar,
       volume = {927},
       number = {1},
          eid = {130},
        pages = {130},
          doi = {10.3847/1538-4357/ac4225},
archivePrefix = {arXiv},
       eprint = {2111.02538},
 primaryClass = {astro-ph.GA},
       adsurl = {https://ui.adsabs.harvard.edu/abs/2022ApJ...927..130P}
}

@ARTICLE{Farber_2022,
       author = {{Farber}, Ryan J. and {Ruszkowski}, Mateusz and {Tonnesen}, Stephanie and {Holguin}, Francisco},
        title = "{Stress-testing cosmic ray physics: the impact of cosmic rays on the surviving disc of ram-pressure-stripped galaxies}",
      journal = {\mnras},
         year = 2022,
        month = jun,
       volume = {512},
       number = {4},
        pages = {5927-5941},
          doi = {10.1093/mnras/stac794},
archivePrefix = {arXiv},
       eprint = {2201.04203},
 primaryClass = {astro-ph.GA},
       adsurl = {https://ui.adsabs.harvard.edu/abs/2022MNRAS.512.5927F}
}

@ARTICLE{Tonnesen_2007,
       author = {{Tonnesen}, Stephanie and {Bryan}, Greg L. and {van Gorkom}, J.~H.},
        title = "{Environmentally Driven Evolution of Simulated Cluster Galaxies}",
      journal = {\apj},
         year = 2007,
        month = dec,
       volume = {671},
       number = {2},
        pages = {1434-1445},
          doi = {10.1086/523034},
archivePrefix = {arXiv},
       eprint = {0709.1720},
 primaryClass = {astro-ph},
       adsurl = {https://ui.adsabs.harvard.edu/abs/2007ApJ...671.1434T}
}

@BOOK{Longair_2011,
       author = {{Longair}, Malcolm S.},
        title = "{High Energy Astrophysics}",
         year = 2011,
       adsurl = {https://ui.adsabs.harvard.edu/abs/2011hea..book.....L}
}

\end{document}